\documentclass[10pt]{article} % For LaTeX2e
\usepackage[preprint]{rlc}
\usepackage{amssymb}            % Defines common symbols like \mathbb R
\usepackage{mathtools}          % Extends amsmath, providing common math tools
\usepackage{mathrsfs}           % Enables \mathscr, which can work in cases that \mathcal does not
\mathtoolsset{showonlyrefs}     % Only number equations that are referenced (optional)
\usepackage{graphicx}           % For including images
\usepackage{subcaption}         % Allows for the use of subfigures and subcaptions
\usepackage[space]{grffile}     % For spaces in image names
\usepackage{url}                % For displaying urls
\usepackage{wrapfig}

\newcommand{\gradientColorEcosystem}{%
    \textcolor{red!10!orange}{C}%
    \textcolor{red!50!orange}{o}%
    \textcolor{magenta!70!red}{l}%
    \textcolor{purple!80!magenta}{o}%
    \textcolor{purple!40!violet!70!magenta}{r}%
    \textcolor{purple!30!violet!70!magenta!80!blue}{E}% 
    \textcolor{purple!60!magenta!50!blue}{c}%
    \textcolor{purple!60!magenta!35!blue}{o}%
    \textcolor{purple!60!magenta!30!blue!80!cyan}{s}%
    \textcolor{purple!60!magenta!30!blue!60!cyan}{y}%
    \textcolor{purple!60!magenta!30!blue!40!cyan}{s}%
    \textcolor{purple!60!magenta!30!blue!20!cyan}{t}%
    \textcolor{blue!40!cyan}{e}%
    \textcolor{blue!60!cyan}{m}%
}

\fancypagestyle{plain}{\pagestyle{fancy}}

\title{\gradientColorEcosystem: Powering Personalized, Standardized, and
Trustworthy Agentic Service in Massive-agent Ecosystem}

\author{ 
\centering
\begin{tabular}{c}
Fangwen Wu\textsuperscript{2}\thanks{Equal contribution.}~~Zheng Wu\textsuperscript{1}\footnotemark[1]~~Jihong Wang\textsuperscript{2}~~Yunku Chen\textsuperscript{2}~~ Ruiguang Pei\textsuperscript{2}\\ 
Heyuan Huang\textsuperscript{2}~~Xin Liao\textsuperscript{2}~~Xingyu Lou\textsuperscript{2}~~Huarong Deng\textsuperscript{2}
\\Zhihui Fu\textsuperscript{2}~~Weiwen Liu\textsuperscript{1}~~Zhuosheng Zhang\textsuperscript{1}~~Weinan Zhang\textsuperscript{1}~~Jun Wang\textsuperscript{2}\thanks{Corresponding author.}
\\
\\[0em]
\textsuperscript{1}Shanghai Jiao Tong University \quad \textsuperscript{2}OPPO
\end{tabular}
}

\begin{document}

\maketitle

\begin{abstract}
With the rapid development of (multimodal) large language model-based agents, the landscape of agentic service management has evolved from single-agent systems to multi-agent systems, and now to massive-agent ecosystems. 
Current massive-agent ecosystems face growing challenges, including impersonal service experiences, a lack of standardization, and untrustworthy behavior.
To address these issues, we propose \textbf{ColorEcosystem}, a novel blueprint designed to enable personalized, standardized, and trustworthy agentic service at scale.
Concretely, ColorEcosystem consists of three key components: agent carrier, agent store, and agent audit. 
The agent carrier provides personalized service experiences by utilizing user-specific data and creating a digital twin, while the agent store serves as a centralized, standardized platform for managing diverse agentic services.
The agent audit, based on the supervision of developer and user activities, ensures the integrity and credibility of both service providers and users.
Through the analysis of challenges, transitional forms, and practical considerations, the ColorEcosystem is poised to power personalized, standardized, and trustworthy agentic service across massive-agent ecosystems.
Meanwhile, we have also implemented part of ColorEcosystem's functionality, and the relevant code is open-sourced at \url{https://github.com/opas-lab/color-ecosystem}.
\end{abstract}

%%%%%%%%%%%%%%%%%%%%%%%%%%%%%%%%%%%%%%%%%%%%%%%%%%%%%%%%%%%%%%%%
%% Section: Submission of papers to RLC
%%%%%%%%%%%%%%%%%%%%%%%%%%%%%%%%%%%%%%%%%%%%%%%%%%%%%%%%%%%%%%%%

%%%%%%%%%%%%%%%%%%%%%%%%%%%%%%%%%%%%%%%%%%%%%%%%%%%%%%%%%%%%%%%%%%%%%%%%

\section{Introduction}
With the development of large language models (LLMs) and multimodal large language models (MLLMs) in key domains such as planning~\citep{valmeekam2023planning,xie2023translating}, reasoning~\citep{wei2022chain,yao2023react}, perception~\citep{huang2023language,zhao2023chat}, and decision-making~\citep{hager2024evaluation,li2022pre}, (M)LLM-based agents have shown strong potential in delivering a wide range of agentic service, such as generating executable code based on user instructions~\citep{zhang2024codeagent,islam2024mapcoder} and directly operating smart terminal devices to fulfill user commands~\citep{zhang2025appagent,wang2024mobile}.

As the scope of agentic service expands, the prevailing paradigm is shifting from isolated single-agent systems to collaborative multi-agent systems, and further toward massive-agent ecosystems.
While massive-agent ecosystems promise substantial growth and diversification of capabilities, they simultaneously introduce critical challenges to system-level coordination and user experience. 
These challenges include:
(i) impersonal or generic user interaction experiences,
(ii) lack of standardization in service management platforms, and
(iii) unreliable or untrustworthy agentic service behavior.

To address these pressing problems in massive-agent ecosystems, we propose \textbf{ColorEcosystem, a novel blueprint designed to enable personalized, standardized, and trustworthy agentic service} at scale. 
ColorEcosystem is built upon three foundational components: agent carrier, agent store, and agent audit.

Within ColorEcosystem, all agentic service activities must be regulated by agent audit. 
ColorEcosystem requires a centralized authoritative entity to conduct agent audits for all developers and users. 
For developers, security audits and information audits are required, while for users, behavior audits and content audits are necessary. 
If any audit fails, the agentic service will be prohibited from execution. 
Under the supervision of the agent audit, the agentic service developed by developers will be uploaded to the agent store for centralized management, making it available for users to invoke.
At the same time, each user will be equipped with an agent carrier. 
The agent carrier contains user-authorized personalized data, which the personalized agent utilizes to make tailored choices when engaging with agentic service. 
Users can select and download the agentic services they need from the agent store into their agent carriers.
%Additionally, the personalized agent can efficiently communicate with other users' personalized agents via the agent protocol.
Additionally, the agent carrier creates digital twins for every users.
The digital twin can efficiently communicate with other users' digital twin via the agent protocol.
This design enables ColorEcosystem to deliver personalized, standardized, and trustworthy agentic service.

As an idealized massive-agent ecosystem, ColorEcosystem still faces numerous challenges and has a long way to go compared to the current chaotic massive-agent ecosystem.  
Therefore, a transitional form of ColorEcosystem could employ graphical user interface (GUI) agents that simulate human operations to fill functional gaps not yet supported by agentic service. 
Additionally, behavior-based supervision can be temporarily used in place of agent audits. 

% We have also conducted practical considerations for ColorEcosystem, analyzing how to attract various agentic service developers to publish their high-quality agentic service in the agent store.

Finally, we provide practical insights into the implementation and adoption of ColorEcosystem, including strategies for incentivizing high-quality agentic service developers to contribute to the agent store.

In the remainder of this paper, we present the details of ColorEcosystem, elaborate on its transitional pathways, and evaluate its practical feasibility, thus offering the vision for the next generation of massive-agent ecosystems.

% In this paper, we propose ColorEcosystem, a novel blueprint for trustworthy, standardized, and personalized massive-agent ecosystems. 

% In the following sections, we will detail ColorEcosystem's core components and transitional forms, and present a practical analysis of its viability, thereby paving the way for the next-generation agent ecosystem.

\section{The Development and Current Problems of Agentic service}
In this section, we first introduce the development of agentic service, from single-agent~\citep{yehudai2025survey} to multi-agent systems~\citep{dorri2018multi} and then to massive-agent ecosystems~\citep{yang2025survey}. 
We then discuss the current problems posed by agentic service at such a massive scale.
\subsection{From a drop, to a wave, to a tsunami: the development of agentic service}
\begin{figure*}[t]
    \centering        
    \includegraphics[width=1\textwidth]{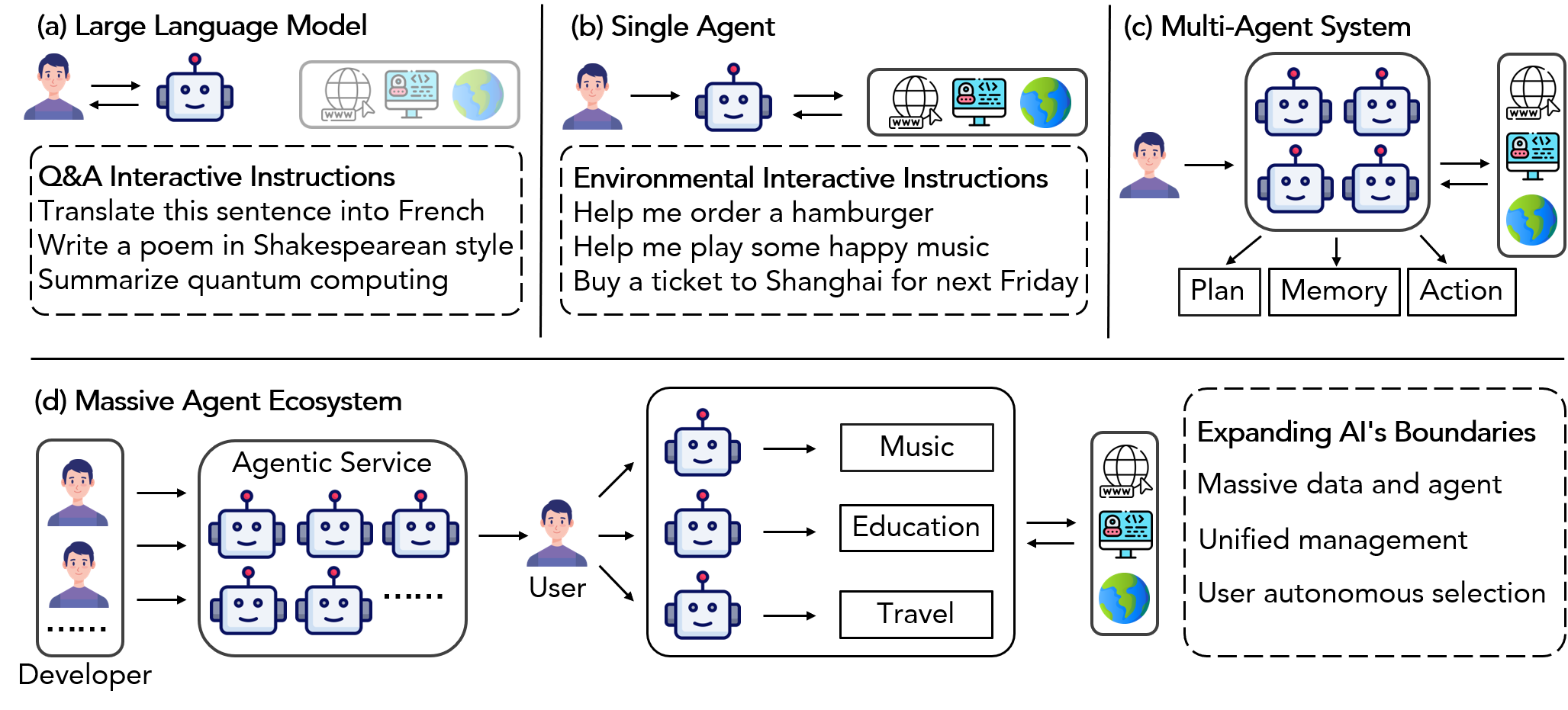} 
    \caption{The development of agentic service. Previously, users interacted with the environment through single-agent or multi-agent systems to accomplish tasks. In the massive-agent ecosystem, developers will create and release massive-agentic service, allowing users to autonomously select different agentic service to accomplish tasks.}
    \label{fig:development} 
\end{figure*}
The development of MLLMs~\citep{huang2024understanding,yao2023react,du2025human,chen2023towards} has spurred the creation of intelligent agents capable of automating complex tasks (e.g., code generation~\citep{zhang2024codeagent,islam2024mapcoder} and operating smart terminal devices~\citep{zhang2025appagent,wang2024mobile}).

As shown in Figure~\ref{fig:development}, the initial paradigm centers on employing an (M)LLM as the central ``brain'' of a single agent, which can interact with the environment through external tools. 
For example, the output of an agent can be regularized into structured formats: Android Debug Bridge commands for operating smartphones~\citep{liu2025llm}; code for libraries like PyAutoGUI to control computers~\citep{hu2024dawn}; or even direct API invocations~\citep{li2025review}.
The single-agent paradigm constructs the initial agentic service.

Single-agent-based agentic service can achieve excellent performance on specific tasks.
However, when confronted with complex problems, a single agent often underperforms due to its inherent limitations. 
Therefore, to leverage the potential of different single agents across various tasks, multiple single agents with distinct functions (e.g., planning, perception, knowledge, memory, reasoning, and evolution) are combined according to specific task requirements to construct multi-agent systems~\citep{li2024survey,han2024llm,maldonado2024multi}. 
This paradigm of multi-agent systems significantly expands the capabilities of agentic service.

With the development of AI agent protocols~\citep{yang2025survey,ehtesham2025survey}, to further expand the capability boundaries of agentic services, developers can publish their agentic service in the form of standardized API interfaces, constrained by the agent protocol, for users to invoke.
The increasing number of participating agentic service will lead to a shift in the paradigm of agentic service towards a massive-agent ecosystem.

\subsection{Navigating the uncharted waters: the current problems of agentic service}

%As shown in Figure~\ref{fig:challenge}, facing the impending massive-agent ecosystem, the paradigm of agentic service confronts three key problems: (i) untrustworthy agentic service behavior, (ii) non-standardized agentic service management platforms, and (iii) impersonal agentic service experiences.

\begin{figure*}[t]
    \centering        
\includegraphics[width=\textwidth]{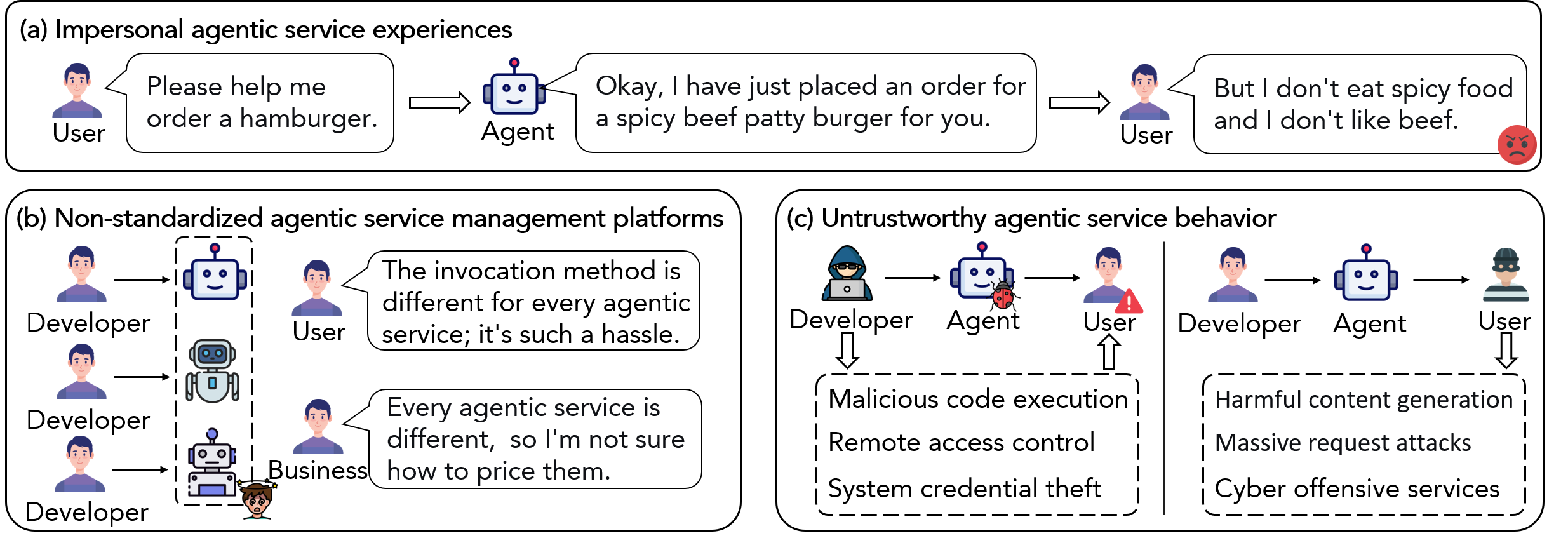} 
    \caption{The current problems of agentic services. (a) Impersonal agent services may fail to align with human intentions. (b) Non-standard agentic service management can cause additional difficulties for user invocation and business billing. (c) Both agentic service developers and users may be malicious.}
    \label{fig:challenge} 
\end{figure*}

As shown in Figure~\ref{fig:challenge}, facing the impending massive-agent ecosystem, the paradigm of agentic service confronts three key problems: (i) impersonal agentic service experiences, (ii) non-standardized agentic service management platforms, and (iii) untrustworthy agentic service behavior.

\subsubsection{Impersonal agentic service experiences}
Prior to the emergence of the massive-agent ecosystem, existing studies~\citep{huang2025advancing,wu2025quick,wang2025perpilot} have recognized the importance of developing personalized agentic service based on users' historical information and preferences.
However, existing agentic services in the massive-agent ecosystem primarily focus on providing unified and convenient service to users, yet to some extent overlook the fact that different users may have different implicit needs, even when given the same instructions.
Impersonal agentic services often carry the risk of misaligning with human true intentions.
For example, a food-ordering agent might order dishes the user dislikes.
Ideally, a set of user-level agents can be established in the massive-agent ecosystem to serve as personalized assistants for each user.

\subsubsection{Non-standardized agentic service management platforms}
Although agent protocols (e.g., MCP~\citep{AnthropicMCP2024} and A2A~\citep{GoogleA2A2025}) standardize the format of agentic service at the function level, there is a lack of platforms that uniformly manage and schedule agentic service developed by different developers using different protocols at the agent level. 
This makes it difficult for users to find the specific agentic service they need among the massive options available.
Meanwhile, for businesses, the lack of unified agentic service management means the inability to establish consistent and balanced pricing standards, which would hinder the formation of an agentic service ecosystem market.
Similar to a smartphone app store, standardized intelligent terminal manufacturers should establish a dedicated store for agentic service to centralize their management and administration.

\subsubsection{Untrustworthy agentic service behavior}
%massive-agentic service bring massive agentic service behavior. 
A massive-agentic ecosystem brings massive agentic service developers and users.
However, on the one hand, not all developers are trustworthy. 
Malicious developers may provide untrustworthy agentic service to carry out attacks (e.g., malicious code execution, remote access control, and credential theft)~\citep{ehtesham2025survey,kumar2025mcp,radosevich2025mcp,cheng2025hidden,wu2025verios}.
On the other hand, not all users are trustworthy. Users might exploit agentic service to perform malicious activities (e.g., spamming, phishing campaigns, or distributed denial-of-service attacks).
Therefore, the agentic service behavior of both developers and users should be subject to appropriate oversight, and the vast number of agentic services in a massive-agent ecosystem poses a significant challenge.

\section{The Blueprint of ColorEcosystem}
\begin{figure*}[h]
    \centering           \includegraphics[width=0.95\textwidth]{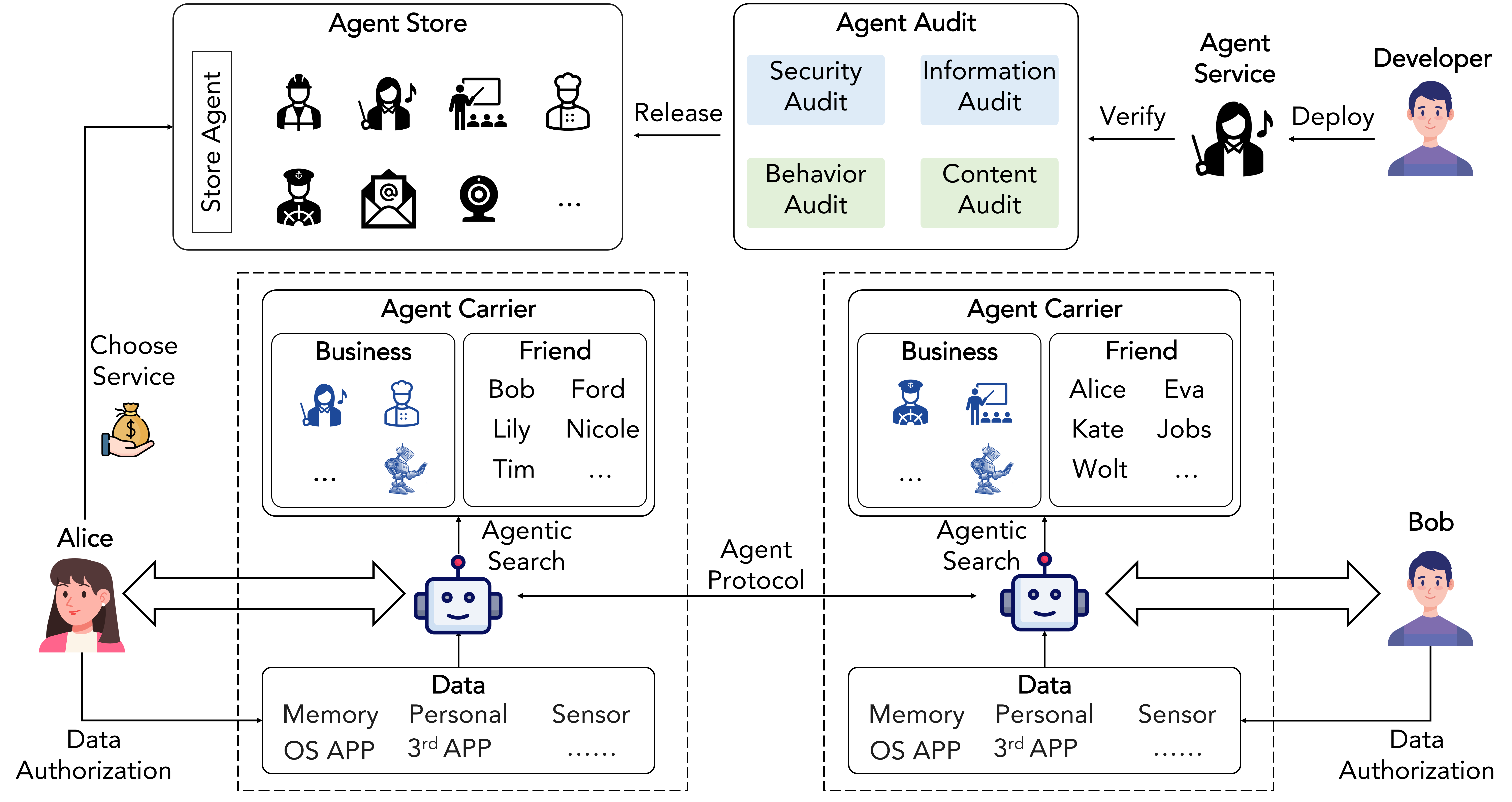} 
    \caption{The Blueprint of ColorEcosystem. ColorEcosystem is a blueprint for a future agentic service ecosystem that ensures trustworthiness through agent audit, while integrating the standard interfaces of the agent store and the user-specific personalized properties of agent carriers.}
    \label{fig:framework} 
\end{figure*}

In response to the current problems of agentic service, we have conceived the blueprint of ColorEcosystem, a massive-agent ecosystem that delivers personalized, standardized, and trustworthy agentic service. Figure~\ref{fig:framework} provides an overview of ColorEcosystem. 
%In this section, we provide an overview of ColorEcosystem, followed by detailed explanations of its three core components: agent audit, agent store, and agent carrier.
In this section, we provide an overview of ColorEcosystem, followed by detailed explanations of its three core components: agent carrier, agent store, and agent audit.

\subsection{The overview of ColorEcosystem}
In ColorEcosystem, human roles are divided into two categories: developers and users.

For developers, once they develop an agentic service, it must first undergo an agent audit. 
The agentic service is then allowed to be published in a standardized format to the agent store.

For users, each user possesses a personalized agent carrier that stores authorized personalized data and memories. 
On one hand, a user Alice’s agent carrier can communicate with a user Bob’s agent carrier by agent protocols to facilitate interaction between users. 
On the other hand, users can select the agentic service they need from the agent store. 
The agent carrier will then personalize the use of these agentic services by incorporating the user’s individual data (e.g., a food-ordering agentic service would take into account the user’s taste preferences during execution).

Moreover, in ColorEcosystem, traditionally defined agents are stored in the agent store in the form of massive-agentic service. 
Under the supervision of the agent audit, these agentic services are continuously published, updated, deleted, and downloaded, similar to mobile apps.

\subsection{Agent carrier}

The design of the agent carrier aims to ensure the personalization of ColorEcosystem.

The agent carrier serves as an important means to transform agents into personalized intelligent assistants, primarily consisting of two core mechanisms.

% First, each user is equipped with a personalized agent. 
% Users authorize a portion of their data to their own agent, which may include their historical trajectories, search records from certain applications, as well as personal information and preferences.

% On one hand, agentic service invoked by the user from the agent store can be executed in a way that better aligns with human intentions through the agent's analysis of the user's data. 
% On the other hand, since agents can communicate with each other using the agent protocol, interactions between users can be simplified into communication between agents.

The first core mechanism is user-specific agentic service selection. 
Users can independently choose the agentic services they need from the agent store and add them to their agent carrier. 
This means the agent carrier does not include all agents from the agent store and only includes those selected by the user. Analogous to an app store, users download apps from the app store, and only the downloaded apps run on their phones.

The second core mechanism is a digital twin capable of accessing user-authorized data. 
Users authorize a portion of their data to their digital twin, which may include their historical trajectories, search records from certain applications, as well as personal information and preferences. 
Agentic services invoked by the user from the agent store can be executed in a way that better aligns with human intentions through the digital twin's analysis of the user's data. 
Since agents can communicate with each other using the agent protocol, interactions between users can be simplified into communication between digital twins.

\begin{figure*}[h]
    \centering        
\includegraphics[width=0.9\textwidth]{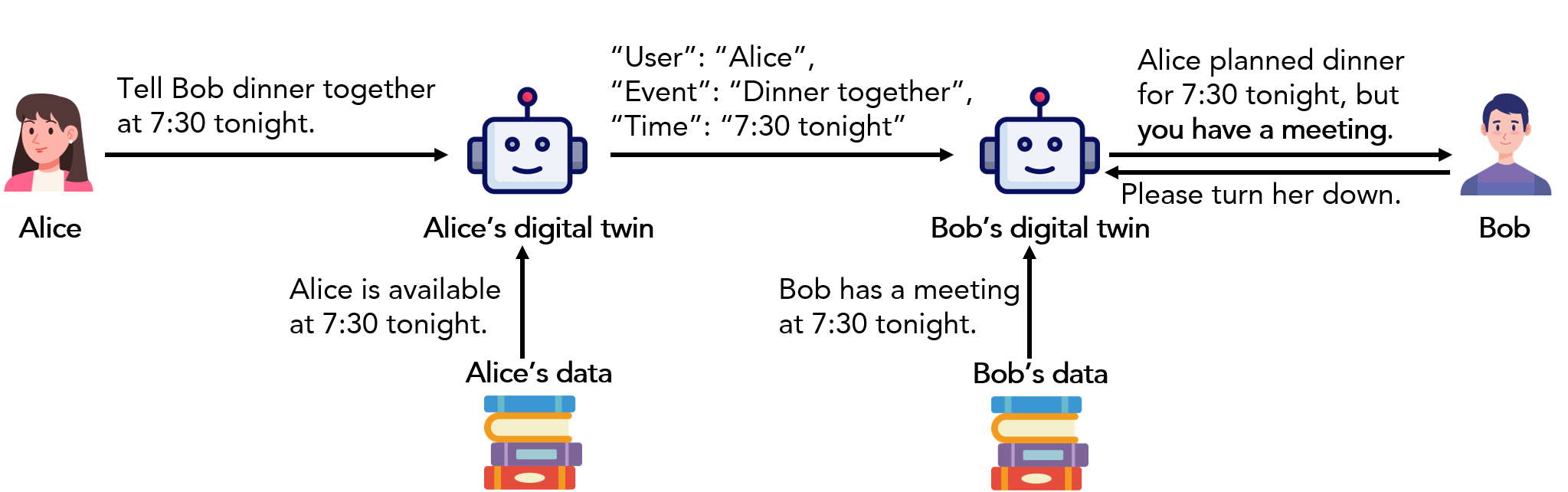} 
    \caption{Example of agent carrier usage. Alice wants to invite Bob to dinner, so Alice's digital twin automatically sends an invitation to Bob's digital twin. Bob's digital twin notifies Bob of the invitation and proactively analyzes Bob's data, informing him that he has a conflicting meeting at the same time.}
    \label{fig:card} 
\end{figure*}

As shown in Figure~\ref{fig:card}, for example, if Alice wants to invite Bob to dinner, she only needs to inform her digital twin. 
Alice's digital twin then analyzes the data authorized by Alice and finds no scheduling conflict. 
It then relays the message to Bob's digital twin in accordance with the agent protocol. 
Bob's digital twin analyzes the data authorized by Bob and detects a conflict between the proposed dinner time and Bob's meeting schedule. 
Therefore, while conveying the dinner invitation to Bob, it also alerts him about the time conflict. 
Such personalized agentic service exemplifies intelligent functionality.

\begin{wrapfigure}{r}{0.5\textwidth}
    \centering
    \includegraphics[width=0.45\textwidth]{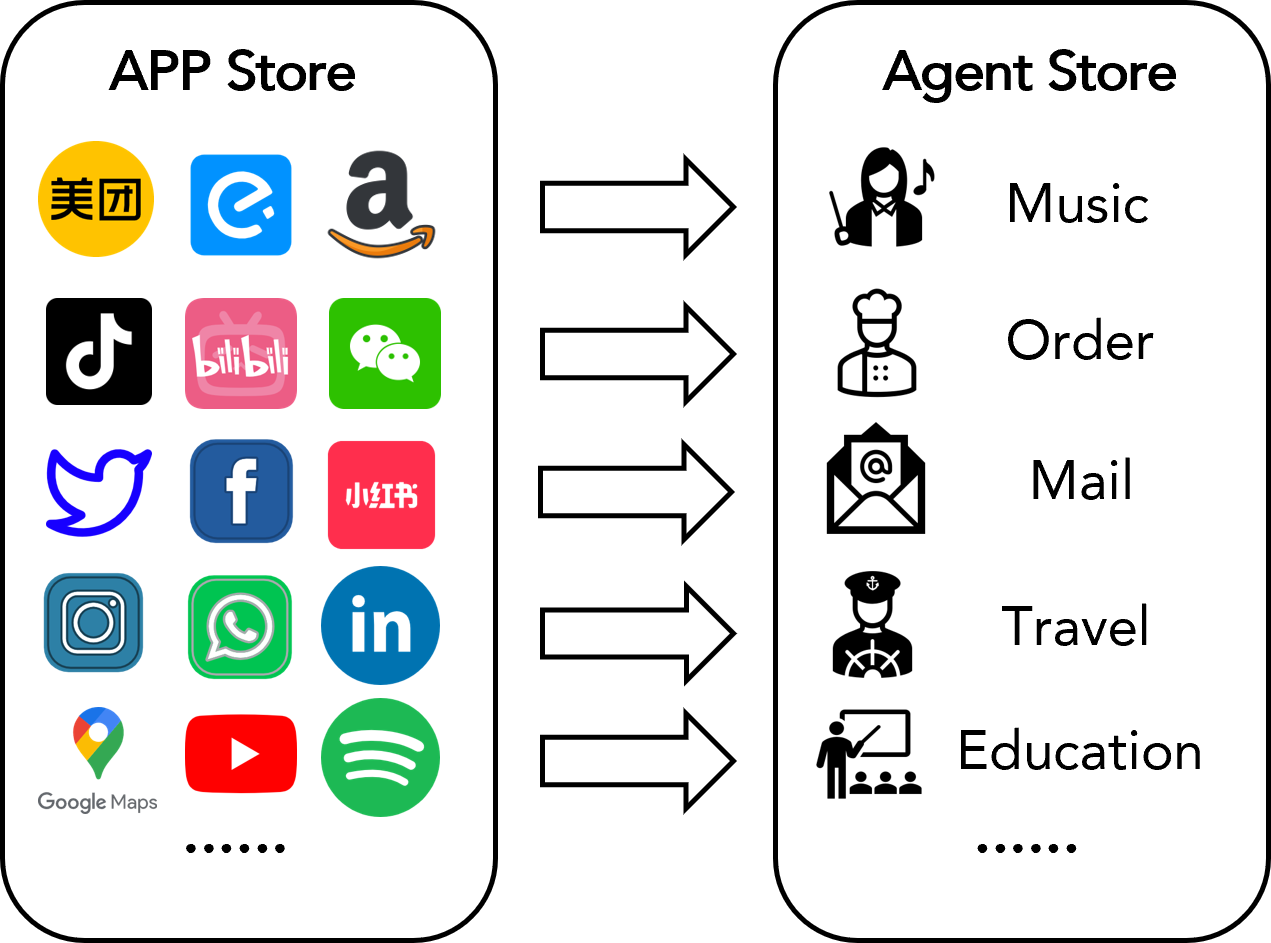}
    \caption{From APP Store to Agent Store: The evolution from individually packaged apps to agentic service encompassing single or multiple functionalities.}
    \label{fig:store}
\end{wrapfigure}

\subsection{Agent store}
The design of the agent store aims to ensure the standardization of ColorEcosystem.

As shown in Figure~\ref{fig:store}, the traditional way developers provide service to users involves creating an application, which users then interact with through a GUI to trigger backend functions.
Various manufacturers have their own APP stores, where developers upload applications for users to download.

In a massive-agent ecosystem, the ideal scenario is that agents can autonomously and automatically complete instructions given by users, making GUI-based applications unnecessary. 
The agent store is a standardized management platform designed for massive-agents. 
Similar to the APP store, developers can release their agentic service to the agent store for users to download and use. 
This approach effectively integrates massive-agentic service resources, enabling users to more conveniently access the most suitable agentic service for their needs.

Compared to current agentic service management platforms (e.g., Coze~\citep{coze}, GPT Store~\citep{gptstore}), the agent store offers greater extensibility. 
Platforms like Coze or GPT Store only allow users to build various AI-inherent agentic services on their platforms based on one or a few base models, meaning the capabilities of these agentic services are constrained by the fixed abilities of the foundation models. 
In contrast, the agent store provides only standardized platforms and management mechanisms, allowing any compliant developers to develop and release their own agentic services on it. 
Developers can use proprietary models from any domain to create agentic services tailored to different tasks.
Meanwhile, the agent store also differs from MuleRun~\citep{mulerun} in that it does not require providing a virtual machine as a runtime environment like MuleRun does. 
Instead, the agent store enables risk-free execution on users' actual devices as the runtime environment through standardized agentic service management.

\subsection{Agent audit}

\begin{figure*}[h]
    \centering          \includegraphics[width=0.95\textwidth]{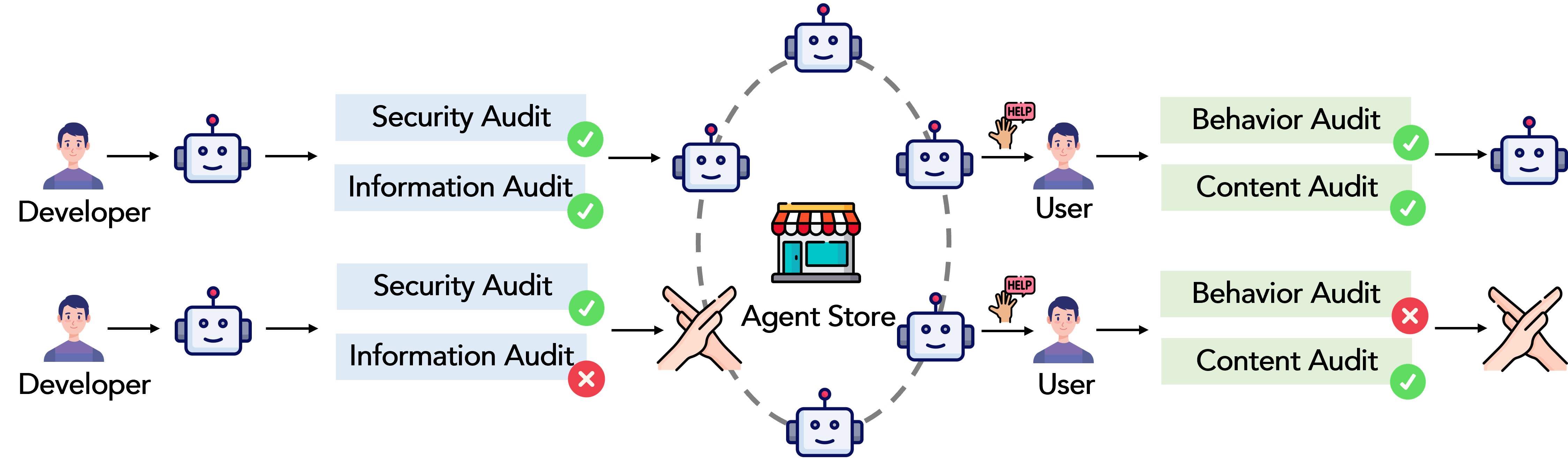} 
    \caption{In ColorEcosystem, developers must undergo security audits and information audits before releasing agentic services to the agent store. Users must undergo behavior audits and content audits before invoking agentic services from the agent store. If any audit fails, the action will be rejected.}
    \label{fig:audit} 
\end{figure*}

The design of the agent audit aims to ensure the trustworthiness of ColorEcosystem.  

As shown in Figure~\ref{fig:audit}, agent audits must be conducted by authoritative third-party institutions with notarization power, and they should be carried out at both the developer and user levels.
If any audit fails, the action will be rejected.

At the developer level, agent audits must include security audits and information audits. 
A security audit refers to examining whether the agentic service contains security risks. 
On one hand, agentic services may have vulnerabilities due to developer oversight; on the other hand, they could be embedded with malicious trojans, viruses, or backdoors by malicious developers. 
Agentic services must fully pass the audit tests under zero-trust conditions before being allowed to be released to the agent store. 
An information audit refers to verifying whether the agentic service includes complete usage instructions and developer information. 
Users need to know how to use the agentic service and should be able to contact the developer for any questions or support issues.

At the user level, agent audits must include behavior audits and content audits.  
A behavior audit examines whether users are exploiting agentic services to engage in malicious activities that could cause harm to other users or the company, thereby ensuring that inherently benign agentic services are not weaponized by malicious users.  
A content audit, on the other hand, ensures that the content generated by users through agentic services does not include defamatory, discriminatory, violent, pornographic, or other offensive material, thus maintaining a healthy and wholesome ecosystem.

% Inspired by the field of blockchain, we believe that agent manufacturers should jointly build a consortium blockchain to record the communication of agentic service on the chain for collective auditing and consensus. 
% This provides an effective method for establishing trustworthiness between manufacturers and manufacturers, as well as between users and manufacturers.

% As shown in Figure~\ref{fig:audit}, in the consortium blockchain~\citep{zheng2018blockchain}, the behaviors of both developers and users need to be collectively audited on the chain. The release of agentic service by developers and the invocation of agentic service by users must obtain consensus from on-chain entities; otherwise, the actions will not be permitted.

% Based on the Byzantine Fault Tolerant theory in blockchain, as long as no more than one-third of the agent manufacturers in ColorEcosystem are malicious~\citep{pease1980reaching}, our designed agent audit can, to some extent, ensure the trustworthiness of ColorEcosystem.

\section{On the Path Toward ColorEcosystem}
In this section, we first introduce the challenges that need to be addressed in realizing ColorEcosystem, followed by proposing a potential transitional form on the path toward ColorEcosystem.

\subsection{Challenges of ColorEcosystem}

Although ColorEcosystem holds promising prospects, the current infrastructure development~\citep{hu2025agents,sager2025comprehensive} still has some distance to go before fully realizing ColorEcosystem. 
This is mainly reflected in the following two aspects: 
\begin{enumerate}
    \item The development of existing agent protocols~\citep{yang2025survey} is still in its early stages and cannot cover every application of each smart terminal device. 
    \item Centralized agent audit requires authoritative bodies or manufacturers within the ecosystem to take the lead, making it difficult to form quickly in a short time.
    \item In the Agent Store, there are massive agentic services, but users may not be able to precisely find the ones that best suit their needs.
\end{enumerate}

Rome was not built in a day. 
Both the functional coverage of agentic service and the establishment of agent audit require cumulative efforts over time. 
Therefore, there is no direct path from the current massive-agent ecosystem to ColorEcosystem.

\subsection{Transitional forms on the path toward ColorEcosystem}

For the above two aspects of challenges, we believe there are transitional compromise solutions.

In cases where existing agentic service cannot cover every application of each smart terminal device, we can deploy GUI agents as an agentic service in the agent store.
A GUI agent is a type of agent that relies on analyzing the current screen and instructions to complete user tasks by simulating human interactions (e.g., clicks, inputs, and swipes).
Since GUI agents do not require access to specific environmental APIs, they possess strong generalization capabilities and can operate any smart terminal device.
Although they may take more time compared to API-based agentic service, they can compensate for the current lack of coverage of agentic service.

As for the difficulty in establishing agent audits, we believe that the only behavior-based regulatory mechanisms already in place within the existing ecosystem, implemented by various manufacturers, can, to some extent, constrain the malicious behaviors of agentic service.
Although behavior-based regulatory mechanisms have an inherent lag, they enable the regulation of massive-agentic service at minimal cost.
On the other hand, although it is currently difficult to establish a centralized agent audit in a short time, a few manufacturers can form a decentralized alliance chain agent audit mechanism as a transitional solution, waiting for the right time to mature and shift to a centralized agent audit.

For how to help users find the most suitable agentic services, the ideal scenario is that the digital twin within the agentic carrier can leverage user data to tailor and assist users in selecting the right agentic services.
A transitional form of this would resemble the push-based recommendation mechanisms used in today’s recommendation systems, where the Agent Store recommends agentic services based on user behaviors such as click-through rates.

\section{Practical Considerations for ColorEcosystem}

As an idealized massive-agent ecosystem, ColorEcosystem requires practical considerations.  
For ColorEcosystem to take shape, developers need to voluntarily and proactively publish agentic service to the agent store.
Agentic service developers can be divided into two categories: transaction-driven and content-driven.  

$\bullet$ Transaction-driven an agentic service developers aim to profit by developing agentic service. 
The payment model for agentic service is typically based on the number of input and output tokens, meaning more user invocations lead to greater profits for transaction-driven agentic service developers.
To achieve more user invocations, these agentic services first need greater exposure.
So ColorEcosystem should follow market principles and user feedback to optimize the recommendation mechanism within the agent store, enabling high-quality transaction-driven developers to generate revenue. 
This will incentivize a broader range of high-quality transaction-driven developers to publish their agentic service to the agent store.  

$\bullet$ As for content-driven developers, who provide high-quality agentic service in the spirit of open source, ColorEcosystem should increase exposure and traffic for their outstanding agentic service in the agent store, encouraging independent developers to publish more of their agentic service.
Additionally, ColorEcosystem should improve the communication mechanism between developers and users, enabling content-driven developers to promptly obtain practical, application-oriented suggestions from user feedback to create higher-quality agent services.

\section{Conclusion}
In this paper, we first conduct a systematic analysis of the development and current problems of agentic service.  
Then, to address the problems posed by the massive-agent ecosystem, we propose a blueprint for ColorEcosystem.  
ColorEcosystem consists of three core components: agent carrier, agent store, and agent audit.  
This design enables ColorEcosystem to deliver personalized, standardized, and trustworthy agentic service.  
Such a paradigm will demonstrate significant value as the massive-agent ecosystem continues to mature.

\section*{Acknowledgments}
We acknowledge the support of the SJTU-OPPO Joint Lab on Artificial Intelligence.

%%%%%%%%%%%%%%%%%%%%%%%%%%%%%%%%%%%%%%%%%%%%%%%%%%%%%%%%%%%%%%%%
%% NOTE: THIS MARKS THE END OF THE "MAIN TEXT"
%%%%%%%%%%%%%%%%%%%%%%%%%%%%%%%%%%%%%%%%%%%%%%%%%%%%%%%%%%%%%%%%

%%%%%%%%%%%%%%%%%%%%%%%%%%%%%%%%%%%%%%%%%%%%%%%%%%%%%%%%%%%%%%%%
%% Bibliography
%%%%%%%%%%%%%%%%%%%%%%%%%%%%%%%%%%%%%%%%%%%%%%%%%%%%%%%%%%%%%%%%
\bibliography{main}
\bibliographystyle{rlc}

\end{document}